\documentclass[conference]{IEEEtran}
\IEEEoverridecommandlockouts

\def\CAMERAREADY{}

\usepackage[pdftex]{graphicx}
\graphicspath{{graphics/}{./}}
\DeclareGraphicsExtensions{.pdf,.jpeg,.png}

\usepackage[utf8]{inputenc} %
\usepackage[T1]{fontenc}    %
\usepackage{url}            %
\usepackage{microtype}      %
\usepackage[dvipsnames]{xcolor}

\usepackage{amsmath,amssymb,amsfonts}
\usepackage{mathtools}
\usepackage{bm}             %
\DeclareMathOperator*{\argmin}{arg\,min}

\usepackage{array}
\usepackage{booktabs}       %
\usepackage{multirow}
\usepackage{threeparttable} %

\usepackage{subcaption}
\usepackage{algorithm}
\usepackage{algpseudocode}  %

\usepackage{csquotes}
\usepackage{xpatch}
\usepackage[%
        backend=biber,                          %
        style=ieee,                             %
        useprefix=true,                         %
    ]{biblatex}
\AtEveryBibitem{\clearfield{note}}
\DeclareFieldFormat{doi}{%
  \mkbibacro{DOI}\addcolon\space
  \ifhyperref
    {\href{https://doi.org/#1}{\nolinkurl{#1}}}
    {\nolinkurl{#1}}}
\DeclareFieldFormat{sentencecase}{#1}
\usepackage[automake, nomain, acronym, xindy, shortcuts=ac]{glossaries-extra}
\glsdisablehyper
\makeglossaries
\setabbreviationstyle[acronym]{long-short} %
\glssetcategoryattribute{acronym}{nohyperfirst}{true}
\newacronym{aru}{ARU}{automatic recording unit}
\newacronym{ssl}{SSL}{self-supervised learning}
\newacronym{bilstm}{BiLSTM}{bidirectional long short-term memory network}
\newacronym{ast}{AST}{audio spectrogram transformer}
\newacronym{vit}{ViT}{vision transformer}
\newacronym{auroc}{AUROC}{area under the receiver operating characteristic curve}
\newacronym{fpr}{FPR}{false positive rate}
\newacronym{tpr}{TPR}{true positive rate}
\newacronym{roc}{ROC}{receiver operating characteristic}
\newacronym{stft}{STFT}{short time Fourier transform}
\newacronym{fft}{FFT}{fast Fourier transform}
\newacronym{nlp}{NLP}{natural language processing}
\newacronym{cnn}{CNN}{convolutional neural network}
\newacronym{ldc}{LDC}{Linguistic Data Consortium}
\newacronym{mfcc}{MFCC}{mel frequency cepstral coefficient}
\newacronym{pcen}{PCEN}{per-channel energy normalisation}
\newacronym{knn}{k-NN}{k-nearest neighbours}
\newacronym{elev}{EV}{Elephant Voices}
\newacronym{elp}{ELP}{Elephant Listening Project}
\newacronym{dt}{DT}{decision tree}
\newacronym{rf}{RF}{random forest}
\newacronym{lr}{LR}{logistic regression}
\newacronym{mlp}{MLP}{multi-layer perceptron}
\newacronym{svm}{SVM}{support vector machine}
\newacronym{xgb}{XGB}{XGBoost}
\newacronym{dft}{DFT}{discrete Fourier transform}
\newacronym{lda}{LDA}{linear discriminant analysis}
\newacronym{lpc}{LPC}{linear predictive coding}
\newacronym{hmm}{HMM}{hidden Markov model}
\newacronym{pam}{PAM}{passive acoustic monitoring}
\newacronym{rnn}{RNN}{recurrent neural network}
\newacronym{map}{mAP}{mean average precision}
\newacronym{ap}{AP}{average precision}
\newacronym{pca}{PCA}{principal component analysis}
\newacronym{resnet}{ResNet}{residual neural network}
\newacronym{iou}{IoU}{intersection over union}
\newacronym{beats}{BEATs}{bidirectional encoder representation from audio transformers}
\newacronym{iucn}{IUCN}{International Union for Conservation of Nature}
\newacronym{rbf}{RBF}{radial basis function}
\newacronym{umap}{UMAP}{uniform manifold approximation and projection}
\newacronym{gpu}{GPU}{graphics processing unit}
\newacronym{lmu}{LMU}{Legendre memory unit}
\newacronym{dec}{DEC}{deep embedded clustering}
\newacronym{nmi}{NMI}{normalised mutual information}
\newacronym{drl}{DRL}{disentangled representation learning}
\newacronym{vae}{VAE}{variational autoencoder}
\newacronym{lstm}{LSTM}{long short-term memory}
\newacronym{gru}{GRU}{gated recurrent unit}
\newacronym{tsne}{t-SNE}{t-distributed stochastic neighbour embedding}
\newacronym{gmm}{GMM}{Gaussian mixture model}
\newacronym{clap}{CLAP}{contrastive language-audio pretraining}
\newacronym{asr}{ASR}{automatic speech recognition}
\newacronym{mds}{MDS}{multidimensional scaling}
\newacronym{dtw}{DTW}{dynamic time warping}
\newacronym{aerd}{AERD}{automatic elephant rumble detection}
\newacronym{dct}{DCT}{discrete cosine transform}
\newacronym{fsl}{FSL}{few-shot learning}

\newglossaryentry{librispeech-960h}{
	name={LibriSpeech ASR~(960h)},
	description={},
}

\newglossaryentry{beans}{
	name={BEANs},
	description={},
}

\newglossaryentry{beans20}{
	name={BEANs-20},
	description={},
}

\newglossaryentry{beans40}{
	name={BEANs-40},
	description={},
}

\newglossaryentry{elman}{
	name={Elman},
	description={},
}

\newglossaryentry{elman-rnn}{
	name={Elman RNN},
	description={},
}

\newglossaryentry{vggish}{
	name={VGGish},
	description={},
}

\newglossaryentry{vgg}{
	name={VGG},
	description={},
}

\newglossaryentry{birdnet}{
	name={BirdNet},
	description={},
}

\newglossaryentry{perch}{
	name={Perch},
	description={},
}

\newglossaryentry{perch8}{
	name={Perch~(ver.~1)},
	description={},
}

\newglossaryentry{surfperch}{
	name={SurfPerch},
	description={},
}

\newglossaryentry{sierras}{
	name={Sierras},
	description={},
}

\newglossaryentry{psla}{
	name={PSLA},
	description={},
}

\newglossaryentry{yamnet}{
	name={YAMNet},
	description={},
}

\newglossaryentry{audiomae}{
	name={AudioMAE},
	description={},
}

\newglossaryentry{hubert}{
	name={HuBERT},
	description={},
}

\newglossaryentry{hubert-base}{
	name={HuBERT~(base)},
	description={},
}

\newglossaryentry{hubert-large}{
	name={HuBERT~(large)},
	description={},
}

\newglossaryentry{hubert-xlarge}{
	name={HuBERT~(xlarge)},
	description={},
}

\newglossaryentry{wav2vec2}{
	name={wav2vec2},
	description={},
}

\newglossaryentry{w2v2-base}{
	name={wav2vec2~(base)},
	description={},
}

\newglossaryentry{w2v2-large}{
	name={wav2vec2~(large)},
	description={},
}

\newglossaryentry{xlsr}{
	name={XLSR},
	description={},
}

\newglossaryentry{multi-whale}{
	name={multi-species whale},
	description={},
}

\newglossaryentry{humpback}{
	name={humpback},
	description={},
}

\newglossaryentry{aves}{
	name={AVES},
	description={},
}

\newglossaryentry{aves-core}{
	name={AVES-core},
	description={},
}

\newglossaryentry{aves-bio}{
	name={AVES-bio},
	description={},
}

\newglossaryentry{aves-nonbio}{
	name={AVES-nonbio},
	description={},
}

\newglossaryentry{aves-all}{
	name={AVES-all},
	description={},
}

\newglossaryentry{birdaves}{
	name={BirdAVES},
	description={},
}

\newglossaryentry{birdaves-biox-base}{
	name={BirdAVES-biox-base},
	description={},
}

\newglossaryentry{birdaves-biox-large}{
	name={BirdAVES-biox-large},
	description={},
}

\newglossaryentry{birdaves-bioxn-large}{
	name={BirdAVES-bioxn-large},
	description={},
}

\newglossaryentry{xenocanto}{
	name={Xeno-canto},
	description={},
}

\newglossaryentry{inaturalist}{
	name={iNaturalist},
	description={},
}

\newglossaryentry{efficientnet-b1}{
	name={EfficientNet-B1},
	description={},
}

\newglossaryentry{birdnetv2.4}{
	name={BirdNET~(ver. 2.4)},
	description={},
}

\newglossaryentry{fsd50k}{
	name={FSD50k},
	description={},
}

\newglossaryentry{audioset}{
	name={AudioSet},
	description={},
}

\newglossaryentry{vggsound}{
	name={VGGSound},
	description={},
}

\newglossaryentry{hdbscan}{
	name={HDBSCAN},
	description={},
}

\newglossaryentry{chatgpt}{
	name={ChatGPT},
	description={},
}

\newglossaryentry{ast-seq}{
	name={AST-seq},
	description={},
}

\newglossaryentry{ast-lab}{
	name={AST-cls},
	description={},
}

\newglossaryentry{lr-frame}{
	name={LR-frame},
	description={},
}

\newglossaryentry{perchv1}{
	name={Perch~(ver.~1)},
	description={},
}

\newglossaryentry{perchv2}{
	name={Perch~(ver.~2)},
	description={},
}

\newglossaryentry{hubert-base-l2}{
	name={HuBERT~(base, layer 2)},
	description={},
}

\newglossaryentry{efficientnet-b0}{
	name={EfficientNet-B0},
	description={},
}

\newglossaryentry{efficientnet-b3}{
	name={EfficientNet-B3},
	description={},
}

\newglossaryentry{wavlm}{
	name={WavLM},
	description={},
}

\newglossaryentry{xeus}{
	name={XEUS},
	description={},
}

\newglossaryentry{proto-net}{
	name={prototypical network},
	description={},
}

\newglossaryentry{relation-net}{
	name={relation network},
	description={},
}

\usepackage{siunitx}
\DeclareSIUnit{\unitless}{\relax}

\usepackage{hyperref}       %
\usepackage{textcomp}
\usepackage{etoolbox}
\usepackage[capitalise, noabbrev, nameinlink]{cleveref}

\usepackage[normalem]{ulem}
\robustify\bfseries
\robustify\uline
\robustify\uuline
\robustify\tnote

\def\BibTeX{{\rm B\kern-.05em{\sc i\kern-.025em b}\kern-.08em
    T\kern-.1667em\lower.7ex\hbox{E}\kern-.125emX}}

\newcommand{\passlr}[1]{\uline{\num{#1}}}
\newcommand{\passrnn}[1]{\uuline{\num{#1}}}

\newcommand{\emailaddr}[1]{{\small\normalfont\texttt{#1}}}
\newcommand{\orcidauthor}[2]{#1\hspace*{1mm}%
  \href{https://orcid.org/#2}{\includegraphics[width=1em]{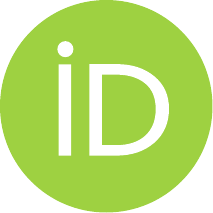}}}

\newcommand{\affilsun}{%
  \textit{Department of Electrical and Electronic Engineering}\\
  \textit{University of Stellenbosch}\\
  Stellenbosch, South Africa}

\newlength{\authorcellwidth}
\newcommand{\authorentry}[3]{%
  \IEEEauthorblockN{\orcidauthor{#1}{#2}}%
  \IEEEauthorblockA{\parbox[t]{\authorcellwidth}{\centering #3}}}

\usepackage[most]{tcolorbox}
\newif\ifreviewmarks
\reviewmarkstrue
\ifdef{\CAMERAREADY}{\reviewmarksfalse}{}

\ifreviewmarks
	\newcommand{\reviewtag}[1]{%
		\textcolor{Red}{\scriptsize \slshape {[REVIEW\ifstrempty{#1}{}{: #1}]}}\space}
\else
	\newcommand{\reviewtag}[1]{}
\fi

\ifreviewmarks
	\newtcolorbox{reviewbox}[1][]{%
		enhanced,
		breakable,
		colback = black!7,
		colframe = Red,
		boxrule = 0.5pt,
		arc = 2pt,
		left = 3pt, right = 3pt, top = 2pt, bottom = 2pt,
		boxsep = 0pt,
		before skip = \smallskipamount,
		after skip = \smallskipamount,
		parbox = false,
		before upper = {\reviewtag{#1}},
	}
\else
	
\fi

\ifreviewmarks
	\AtBeginDocument{%
		\setlength{\textfloatsep}{7pt plus 2pt minus 2pt}%
		\setlength{\dbltextfloatsep}{7pt plus 2pt minus 2pt}%
		\setlength{\floatsep}{6pt plus 2pt minus 2pt}%
		\setlength{\dblfloatsep}{6pt plus 2pt minus 2pt}%
		\setlength{\intextsep}{6pt plus 2pt minus 2pt}%
		\setlength{\abovedisplayskip}{4pt plus 2pt minus 2pt}%
		\setlength{\belowdisplayskip}{4pt plus 2pt minus 2pt}%
		\setlength{\abovedisplayshortskip}{2pt plus 1pt minus 1pt}%
		\setlength{\belowdisplayshortskip}{3pt plus 1pt minus 1pt}%
	}
\fi

\ifdef{\CAMERAREADY}{%
	\AtBeginDocument{%
		\setlength{\textfloatsep}{6pt plus 2pt minus 2pt}%
		\setlength{\dbltextfloatsep}{6pt plus 2pt minus 2pt}%
		\setlength{\floatsep}{6pt plus 2pt minus 2pt}%
		\setlength{\dblfloatsep}{6pt plus 2pt minus 2pt}%
		\setlength{\intextsep}{6pt plus 2pt minus 2pt}%
		\setlength{\abovedisplayskip}{5pt plus 1pt minus 1pt}%
		\setlength{\belowdisplayskip}{5pt plus 1pt minus 1pt}%
		\setlength{\abovedisplayshortskip}{2pt plus 1pt minus 1pt}%
		\setlength{\belowdisplayshortskip}{2pt plus 1pt minus 1pt}%
	}%
	\captionsetup{skip=5pt}%
}{}

\newtoggle{blindreview}
\toggletrue{blindreview}
\ifdef{\CAMERAREADY}{\togglefalse{blindreview}}{}
\iftoggle{blindreview}
  {\typeout{BLIND mode: review copy}}
  {\typeout{BLIND mode: camera-ready}}

\usepackage[switch]{lineno}

\newsavebox{\hiddenbox}
\newlength{\hiddenwd}
\newlength{\hiddenht}
\newlength{\hiddendp}
\newlength{\hiddenindent}
\newenvironment{hidden}[1][\textit{Withheld for purpose of blind peer-review}]{%
  \def\hiddennotice{#1}%
  \setlength{\hiddenindent}{\parindent}%
  \begin{lrbox}{\hiddenbox}%
  \begin{minipage}[t]{\linewidth}\setlength{\parindent}{\hiddenindent}%
}{%
  \end{minipage}%
  \end{lrbox}%
  \setlength{\hiddenwd}{\wd\hiddenbox}%
  \setlength{\hiddenht}{\ht\hiddenbox}%
  \setlength{\hiddendp}{\dp\hiddenbox}%
  \iftoggle{blindreview}{%
    \sbox{\hiddenbox}{\parbox[t]{\hiddenwd}{\centering\hiddennotice}}%
    \ifdim\ht\hiddenbox<\hiddenht \ht\hiddenbox\hiddenht \fi
    \ifdim\dp\hiddenbox<\hiddendp \dp\hiddenbox\hiddendp \fi
    \ifdim\dimexpr\ht\hiddenbox+\dp\hiddenbox\relax
         >\dimexpr\hiddenht+\hiddendp\relax
      \PackageWarning{blindreview}{Notice does not fit the space it replaces}%
    \fi
  }{}%
  \par\noindent\usebox{\hiddenbox}\par
}

\makeatletter
\newsavebox{\blindtitlea}
\newsavebox{\blindtitleb}
\newlength{\blinddelta}

\newcommand{\blindmeasure}[2]{%
  \global\setbox#1\vbox{%
    \hsize\textwidth \@parboxrestore \col@number\@ne
    \normalfont\normalsize
    \hbadness\@M \vbadness\@M \hfuzz\maxdimen \vfuzz\maxdimen
    \IEEEquantizedisabletitlecmds
    \let\@author#2%
    \@maketitle}%
}

\newcommand{\reserveauthorspace}{%
  \blindmeasure\blindtitlea\authorreal
  \blindmeasure\blindtitleb\authorblind
  \setlength{\blinddelta}{\dimexpr\ht\blindtitlea-\ht\blindtitleb\relax}%
  \ifdim\blinddelta<\z@
    \PackageWarning{blindreview}{Notice is taller than the author block}%
  \else
    \setlength{\blindpadabove}{0.5\blinddelta}%
    \setlength{\blindpadbelow}{\dimexpr\blinddelta-\blindpadabove\relax}%
    \blindmeasure\blindtitleb\authorblind
    \addtolength{\blindpadbelow}{\dimexpr\ht\blindtitlea-\ht\blindtitleb\relax}%
  \fi
  \blindmeasure\blindtitleb\authorblind
  \typeout{BLIND title height: real=\the\ht\blindtitlea\space
    blind=\the\ht\blindtitleb\space reserved=\the\blinddelta}%
  \ifdim\ht\blindtitlea=\ht\blindtitleb\else
    \PackageWarning{blindreview}{Title height mismatch}%
  \fi
}
\makeatother

\begin{document}

\iftoggle{blindreview}{\linenumbers}{}

\title{A Parameter-Free Few-Shot Evaluation for\\Elephant Vocalisation Classification}

\newcommand{\authorreal}{%
\authorentry{Christiaan M. Geldenhuys}{0000-0003-0691-0235}{%
  \affilsun\\\emailaddr{cmgeldenhuys@sun.ac.za}}%
\and%
\authorentry{Thomas R. Niesler}{0000-0002-7341-1017}{%
  \affilsun\\\emailaddr{trn@sun.ac.za}}%
}

\newlength{\blindpadabove}
\newlength{\blindpadbelow}
\newcommand{\authorblind}{%
  \IEEEauthorblockN{\\[\blindpadabove]%
    \textbf{\texttt{Authors omitted for purpose of blind peer-review}}%
    \\[\blindpadbelow]}%
}

\author{\authorreal}
\iftoggle{blindreview}{\reserveauthorspace\author{\authorblind}}{}

\hypersetup{
hidelinks,
pdftitle={A Parameter-Free Few-Shot Evaluation for Elephant Vocalisation Classification},
pdfkeywords={few-shot learning, nearest-centroid classification, bioacoustics, elephant vocalisations, pretrained embeddings, Perch},
}
\iftoggle{blindreview}{}{%
  \hypersetup{pdfauthor={Christiaan M.~Geldenhuys, Thomas R.~Niesler}}}
\maketitle

\begin{abstract}
	We present a \emph{parameter-free} episodic evaluation of nearest-centroid classification of elephant vocalisations on fixed pretrained embeddings, for the \ac{elev} and \ac{ldc} datasets.
	We ask not which embedding yields the best classifier trained on all labelled data, but how the simplest classifier performs as the number of exemplars per class varies.
	There are \emph{no learnable parameters}, because each class is modelled as the mean of its support embeddings and each query is assigned to the nearest centroid under squared Euclidean distance.
	Evaluation covers the fixed \gls{perchv1}, \gls{perchv2} and \gls{hubert-base-l2} embeddings, alongside \ac{mfcc} features, $N$-way $k$-shot, under the same stratified $K$-fold cross-validation protocol as the trained classifiers.
	None of these embedding models was trained to distinguish elephant call types.
	On the smaller \ac{elev} dataset the centroid classifier is markedly data-efficient.
	Using \gls{perchv1} or \gls{perchv2} embeddings it overtakes in \ac{map} the fully-trained \ac{lr} baseline from one or two exemplars and the recurrent baseline from two.
	Over the reduced set of call types on which the strongly-supervised end-to-end baseline was trained, the centroid classifier using \gls{perchv2} embeddings overtakes that baseline in \ac{map} as well, from two exemplars.
	On the larger \ac{ldc} dataset the recurrent baselines retain their advantage for all considered values of $k$.
	Only \ac{lr} is overtaken, and only in \ac{map}.
	Nearest-centroid classification is therefore preferable precisely when exemplars are few and the fixed embedding already separates the call types.
\end{abstract}


\section{Introduction}
\label{sec:intro}

The African bush elephant (\textit{Loxodonta africana}) and the Asian elephant (\textit{Elephas maximus}) are both classified as endangered~\autocite{iucn2020loxodontaafricana, iucn2019elephasmaximus}.
Automated classification of elephant vocalisations advances the study of elephant behaviour, population distribution and conservation requirements~\autocite{zeppelzauer2015eledetsystem}.
Such systems support continuous passive monitoring and inform wildlife management in protected areas and conservation facilities.

Although automated elephant call classification has been attempted~\autocite{clemins2003elepspeech,zeppelzauer2015autoclass}, the scarcity of labelled data has limited the application of deeper neural network approaches and has, to date, prevented substantial performance gains~\autocite{bjorck2019elepcnn,geldenhuys2024aerd}.
Obtaining expertly labelled recordings is expensive, since correctly identifying and labelling vocalisations requires domain experts in bioacoustics or ecology.
Even under strict annotation conventions, labels may be noisy.
Common causes include disagreement on annotated start and end times, calls obscured by interfering sources, overlapping vocalisations and inter-annotator variability.

Annotation, rather than computation, remains a dominant cost in conservation bioacoustics.
Since the observed behaviour forms part of the subcall definition, and this generally requires field observation rather than audio alone, very few exemplars are available per subcall type.
Rather than asking which embedding model yields the best classifier when trained on \emph{all} available labelled data, we therefore ask how well the simplest possible classifier performs on fixed pretrained embeddings as the number of labelled exemplars per class is varied.
None of the embedding models we consider was trained to distinguish elephant call types, so the evaluation also measures how well representations learnt from other audio transfer to this task.
The classifier we consider represents each class by a centroid, the mean of its support embeddings, and assigns each query to the nearest centroid.
The key advantage of this approach is that it has \emph{no learnable parameters}, which makes it feasible for very low-resource datasets~(five or fewer exemplars per class).
The value of this simplicity depends on how scarce the labelled exemplars are and on how well the fixed pretrained embedding already separates the classes, and the two datasets we evaluate span the resource scale from very low to abundant.

\subsection{Contributions}
\label{sec:contributions}

To the best of our knowledge, this work makes the following novel contributions.
\begin{itemize}
	\item
	      We present the first episodic few-shot evaluation of elephant call and subcall classification, using a classifier with no learnable parameters.

	\item
	      We present a comparison of this classifier against an end-to-end system trained for the task and against trained classifiers using the same fixed embeddings, establishing which is preferable given the quantity of labelled data available.

	\item
	      An analysis, by nonparametric bootstrap, of the sampling noise that randomly drawn support sets introduce into \ac{fsl}, which reports classification performance as a 1st-to-99th-percentile interval rather than as a single point estimate.

	\item
	      An analysis of the effect that an unequal number of labelled exemplars per class has on \ac{fsl}.
		  This is relevant when the pool of labelled examples of certain classes is exhausted.
\end{itemize}

\section{Background}
\label{sec:background}

Three fields converge in this work, namely elephant call classification, acoustic embedding models and few-shot learning, the last of which is treated separately in \cref{sec:background:fsl}.
Deep learning in bioacoustics is reviewed by \textcite{stowell2022biodeep}, and pretrained embedding models for bioacoustic classification by \textcite{Geldenhuys2026Embed}.

\subsection{Elephant call classification}
\label{sec:background:elephant}

Elephant rumbles, which are low-frequency vocalisations, convey information such as age, sex, reproductive status and emotional state~\autocite{poole1988social}, so understanding vocal behaviour is central to conservation and socio-ecological research~\autocite{zeppelzauer2015autoclass}.
Computational work on elephant vocalisations spans more than two decades, from automated detection of free-ranging rumbles in field recordings~\autocite{zeppelzauer2015eledetsystem} to \ac{cnn} detectors with differentiable compression that address the bandwidth constraints of remote \ac{pam} deployments~\autocite{bjorck2019elepcnn}.
More recently, \textcite{Pickering2025TransferELPBehaviour} fine-tuned pretrained \ac{cnn} embeddings for forest elephant call classification, demonstrating that general-purpose audio embeddings transfer effectively to elephant bioacoustics.

Labelled recordings nonetheless remain scarce, which has limited the application of more recent neural network approaches to elephant call classification~\autocite{clemins2003elepspeech,zeppelzauer2015autoclass,geldenhuys2024aerd}.
An evaluation of pretrained embedding models on fully labelled data indicates that, while such models are more data-efficient than training deep networks from random initialisation, a parametric classifier must still be trained for them to perform well.
That classifier requires labelled data for both training and model selection, and adds a hyperparameter-tuning burden which is not feasible when labelled exemplars are very scarce~\autocite{Geldenhuys2026Embed}.

\subsection{Acoustic embedding models}
\label{sec:background:embedding}

Acoustic embedding models are deep neural networks trained to represent an audio signal as a fixed, high-dimensional numerical vector, referred to here as an acoustic embedding, or simply an embedding.
Such models are typically trained on very large unlabelled audio corpora by self-supervision, discovering patterns in the data without the aid of human annotations.
Because the data are unlabelled, the risk of annotator bias is reduced, although these approaches require substantially larger datasets and greater computation than their supervised counterparts.
Once trained, such models have been shown to generalise to out-of-domain tasks, including taxa and recording conditions absent from the pretraining data~\autocite{Burns2025Perch2Whales}.

\section{Few-Shot Learning}
\label{sec:background:fsl}

\Ac{fsl} is the task of classifying new inputs given only a small number of labelled exemplars per class.
When performing conventional supervised training, adding a class requires the collection of many further annotated samples and then retraining the classifier.
This is particularly limiting where annotation is costly or the recordings available per class are inherently few.
In contrast, \acf{fsl} structures training and evaluation so that classifiers must generalise from a minimal set of labelled exemplars.
\Textcite{Nolasco2023BioFSL} formalised the DCASE few-shot bioacoustic detection challenge, in which prototypical networks detect animal sounds from as few as five labelled examples.

One approach to \ac{fsl} is to use a pretrained embedding model as a fixed feature extractor and apply a shallow or similarity-based classifier to its output representations~\autocite{Geldenhuys2026Embed}.
\Textcite{mcewen2023asthumaninloop} applies a prototypical network on fixed pretrained embeddings to detect rare animal vocalisations.
This represents the closest prior work to the evaluation presented here.
Few-shot classification using fixed pretrained embeddings has thus been applied to rare animal vocalisations and to whale calls~\autocite{mcewen2023asthumaninloop,Burns2025Perch2Whales}, and pretrained embeddings have been fine-tuned for forest elephant call classification~\autocite{Pickering2025TransferELPBehaviour}.
To the best of our knowledge, however, no few-shot evaluation of elephant call or subcall classification has yet been reported.

\subsection{Episodic evaluation}
\label{sec:background:fsl:episodic}

Few-shot classifiers are evaluated \emph{episodically}, following the protocol of \textcite{Vinyals2016MatchingNetworks}.
An \emph{episode}~$\mathcal{E}$ comprises a support set and a disjoint query set.
The \emph{support set}~$\mathcal{S}$ holds the labelled exemplars on which the classifier grounds its decisions.
The \emph{query set}~$\mathcal{Q}$ holds the examples to be classified, and their labels are withheld and used only to score the resulting classifications.
Keeping the two sets disjoint ensures that no input~(query) is classified using itself as evidence~(support).

An episode is defined by two numbers, $N$ and $k$, where $N$ is the number of classes among which each query must be classified and $k$ is the number of labelled exemplars provided per class.
Of the $C$ classes in the dataset, $N \le C$ are drawn at random, and $k$ labelled exemplars are sampled for each \emph{without replacement}.%
\footnote{A class holding fewer than $k$ labelled exemplars instead contributes its entire eligible pool~(see \cref{sec:setup:partial}).}
This is referred to as an $N$-way $k$-shot classification task, whose support set is formed from $m = N \times k$ audio segments $\mathbf{x}_i$ with associated annotated labels $y_i$,
\begin{equation}
	\mathcal{S} = \{(\mathbf{x}_i,\, y_i)\}_{i=1}^{m}, \quad m = N \times k.
	\label{eq:support}
\end{equation}
A further $q$ examples, disjoint from $\mathcal{S}$ and drawn \emph{without replacement}, form the query set~$\mathcal{Q}$,
\begin{equation}
	\mathcal{Q} = \{(\mathbf{x}_j,\, \underbrace{y_j}_{\mathclap{\text{withheld}}})\}_{j=1}^{q}.
	\label{eq:query}
\end{equation}

\begin{figure}[tb]
	\centering
	\includegraphics[width=\columnwidth]{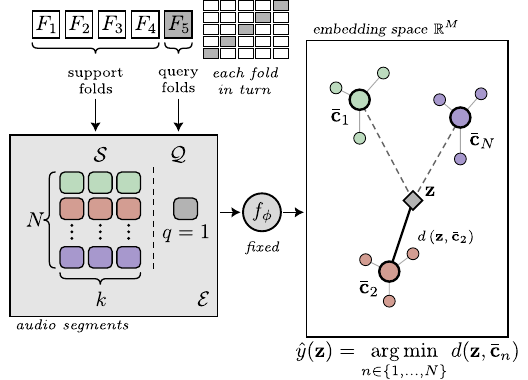}
	\caption{%
		Episode construction and the centroid classifier.
		Each segment is mapped by the fixed embedding function~$f_\phi$ to an embedding.
		Support embeddings, one colour per class, are averaged into class centroids, and a query is assigned to the nearest centroid.
		The cells $F_1$ to $F_5$ mark the stratified cross-validation folds, one held out as the query fold and the remainder forming the support folds from which the support set is drawn.
		Each fold serves as the query fold in turn~(see \cref{sec:setup:eval}).
	}
	\label{fig:method}
\end{figure}

Episodes are conventionally compiled in succession for \emph{episodic training}, under a protocol that partitions the dataset at the \emph{class level} so that the classes seen at evaluation are disjoint from those seen during training~\autocite{Vinyals2016MatchingNetworks}.
The centroid classifier, however, contains no learnable parameters, and the embedding models are pretrained~(see \cref{sec:emb-model}).
No training or development sets are therefore required, and we apply the episodic framework to evaluation only.
It is here that we depart from the conventional protocol, setting $N = C$ so that every class occurs in every episode, partitioning the data into recording-grouped cross-validation folds rather than into disjoint subsets of classes, and composing each episode as set out in~\cref{sec:setup:eval}.
\Cref{fig:method} illustrates the construction of an episode.

\subsection{Nearest-centroid classifier}
\label{sec:background:centroid}

The \emph{nearest-centroid} classifier is a parameter-free method with a long history in information retrieval~\autocite{Rocchio1971Retrieval} and statistical learning~\autocite{Hastie2009ProtoMethodNearestNeighbors}.
We state it here in terms of the support and query sets of an episode.
An embedding function $f_\phi$, with pretrained parameters $\phi$ held fixed throughout, maps each support segment $\mathbf{x}_i$ to an $M$-dimensional embedding $\mathbf{z}_i = f_\phi(\mathbf{x}_i) \in \mathbb{R}^{M}$.
The centroid for class $n$ is the arithmetic mean of its support embeddings,
\begin{equation}
	\bar{\mathbf{c}}_n = \frac{1}{\lvert \mathcal{S}_n \rvert} \sum_{(\mathbf{x}_i,\, y_i) \in \mathcal{S}_n} \mathbf{z}_i,
	\label{eq:centroid}
\end{equation}
where $\mathcal{S}_n \subseteq \mathcal{S}$ holds the support examples of class~$n$.
A query embedding $\mathbf{z} = f_\phi(\mathbf{x})$ is given the class $\hat{y}(\mathbf{z})$ of the nearest centroid among the $N$ classes of the episode, under a pairwise dissimilarity function $d(\cdot,\cdot)$,
\begin{equation}
	\hat{y}(\mathbf{z}) = \argmin_{n \in \{1, \ldots, N\}} d\!\left(\mathbf{z}, \bar{\mathbf{c}}_n\right).
	\label{eq:centroid:argmin}
\end{equation}

\paragraph{Squared Euclidean distance}
Throughout this work $d(\cdot, \cdot)$ is the squared Euclidean distance,
\begin{equation}
	d\!\left(\mathbf{z}, \bar{\mathbf{c}}_n\right) = \lVert \mathbf{z} - \bar{\mathbf{c}}_n \rVert^{2},
	\label{eq:centroid:sqeuclid}
\end{equation}
which is a Bregman divergence, and for such divergences the arithmetic mean uniquely minimises the total within-class distance~\autocite{Banerjee2005Bregman}.
Defining each class representative as the mean of~\cref{eq:centroid} is therefore the principled choice, and is also the rule underlying prototypical networks~\autocite{Snell2017Protonets}.

\paragraph{Continuous per-class scores}
The assignment in~\cref{eq:centroid:argmin} is hard, whereas the evaluation calls for a continuous per-class score~(see \cref{sec:metrics}).
Following \textcite{Snell2017Protonets}, passing the negative distances through a softmax gives
\begin{equation}
	s_n(\mathbf{z}) = \frac{\exp\!\left(-d(\mathbf{z},\, \bar{\mathbf{c}}_n)\right)}{\sum_{n'=1}^{N} \exp\!\left(-d(\mathbf{z},\, \bar{\mathbf{c}}_{n'})\right)}, \qquad n = 1, \ldots, N,
	\label{eq:centroid:softmax}
\end{equation}
a score vector over the $N$ classes.
The softmax is monotonic, so the class of greatest score is that of~\cref{eq:centroid:argmin}.
The conversion introduces no temperature parameter, leaving the classifier free of hyperparameters.

A new centroid is formed from the support set in every evaluation episode, and no state is carried between episodes.
No weights are trained, no validation set is required, no early stopping is applied and no hyperparameter is tuned.
No data therefore need be reserved for training or development, and the whole dataset is evaluated by cross-validation~(see \cref{sec:setup:eval}).
The burden of discrimination falls entirely on the fixed pretrained embedding.

\paragraph{Centroid classifier as fixed linear layer}
Expanding \cref{eq:centroid:sqeuclid} and discarding the $\lVert \mathbf{z} \rVert^{2}$ term, constant in $n$, reduces the assignment rule to a per-class linear form,
\begin{gather}
	\hat{y}(\mathbf{z}) = \argmin_{n \in \{1, \ldots, N\}} \underbrace{\left( \mathbf{w}_n^\top \mathbf{z} + b_n \right)}_{\mathclap{\text{per-class linear layer}}},
	\label{eq:centroid:linear} \\
	\mathbf{w}_n = -2\bar{\mathbf{c}}_n, \quad b_n = \lVert \bar{\mathbf{c}}_n \rVert^{2}, \notag
\end{gather}
whose weights and biases are set entirely by the class centroids.
With the softmax of~\cref{eq:centroid:softmax}, this is a multinomial logistic regression with fixed rather than fitted weights.

Fixing the weights to the centroids is not a neutral choice.
It encodes an isotropic within-class covariance, an inductive bias rather than a property of any particular embedding.
Estimating a full covariance instead would need a number of exemplars growing with the embedding dimension $M$, far more than the few per class available here~\autocite{Marchenko1967EigenCovar,Ledoit2004LargeCovarEst}.
The advantage of fixing the weights is therefore greatest when $k$ is smallest, which is where this work concentrates.

\section{Datasets}
\label{sec:data}

We use two datasets, the \ac{elev} and \ac{ldc} datasets, evaluated separately because they differ in size, recording setup and species.
Both are summarised in \cref{tab:data}.

\begin{table}[tb]
	\centering
	\begin{threeparttable}
		\caption{
			The \ac{elev} and \ac{ldc} datasets, and the subset of each that is evaluated.
			The upper block describes the long-form recordings as captured, and the lower block the segments derived from them~(see \cref{sec:segmentation}).
			Call types are counted at the annotation level of each dataset, so the two are not directly comparable.
			The shot count $k^\star$ is defined in \cref{sec:setup:partial}.
		}
		\label{tab:data}
		\small
		\setlength{\tabcolsep}{2pt}
		\begin{tabular}{@{}rcc@{}}
			\toprule
			Dataset                & \textit{\acs{elev}}         & \textit{\acs{ldc}}              \\ \midrule
			Authors                & \textcite{poole2021elev}    & \textcite{ldc2010asianelevoc}   \\
			Elephant species       & \textit{Loxodonta africana} & \textit{Elephas maximus}        \\
			Sampling rate          & \qty{44.1}{\kilo\hertz}     & \qty{48}{\kilo\hertz}           \\
			Detail of annotation   & Recording-level             & Within \qty{100}{\milli\second} \\
			Taxonomic level        & Subcall type                & Call type                       \\
			Call types annotated   & 27                          & 16                              \\
			Number of recordings   & 226                         & 1577                            \\
			Total duration         & \qty{55.6}{minutes}         & \qty{57.4}{hours}               \\
			Annotated duration     & \qty{55.6}{minutes}         & \qty{32.6}{hours}               \\
			Cross-validation folds & 5                           & 10                              \\ \midrule
			Annotated segments               & 422                 & 4845                \\
			Number of call types, $C$        & 12                  & 10                  \\
			Query segments, $Q$              & 257                 & 3985                \\
			Evaluated duration               & \qty{17.9}{minutes} & \qty{5.37}{hours}   \\
			Average segment length           & \qty{4.16}{\second} & \qty{4.85}{\second} \\
			Complete shot count, $k^\star$   & 4                   & 31                  \\ \bottomrule
		\end{tabular}
	\end{threeparttable}
\end{table}

The \ac{elev} dataset comprises field recordings of free-roaming African bush elephants~(\textit{Loxodonta africana}) made by~\citeauthor{poole2021elev} at the non-profit organisation Elephant Voices~\autocite{poole2021elev}.
Annotations are at the recording level and give no within-recording call timings, so recordings with more than one call were divided manually into single-call segments.
Where the annotation is specific enough, a segment also carries a subcall type from the taxonomy of~\textcite{poole2021elev}.

The \ac{ldc} dataset comprises field recordings of free-roaming Asian elephants~(\textit{Elephas maximus}) made by~\citeauthor{ldc2010asianelevoc} and hosted by the \ac{ldc}~\autocite{ldc2010asianelevoc}, annotated according to their call taxonomy.
Its 16 categories comprise 14 vocal call types and two non-vocal signals, the \textit{blow} and the \textit{trunk-bounce}.
The annotations give start and end times within a long-form recording, so segments are cut directly, with a \qty{250}{\milli\second} collar at each end.

\begin{figure}[tb]
	\centering
	\includegraphics[width=\columnwidth]{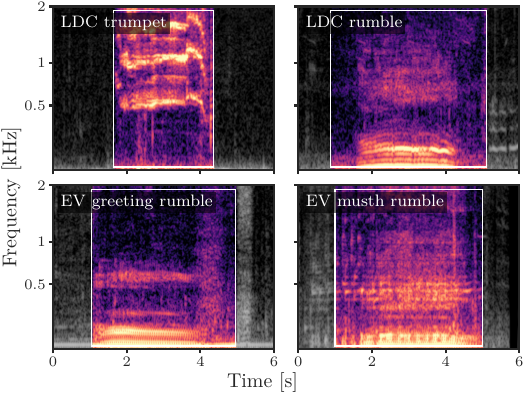}
	\caption{%
		Log-mel spectrograms of four vocalisations.
		Each bounding box spans the annotated call.
		The upper row shows two acoustically distinct \ac{ldc} call types, the lower row two similar \ac{elev} subcall types.
	}
	\label{fig:calltypes}
\end{figure}

The two corpora are thus annotated at different levels of the call taxonomy.
The \ac{elev} annotations distinguish fine-grained subcall types such as the \textit{greeting rumble} and the \textit{musth rumble}, whereas the \ac{ldc} annotations distinguish coarser call types such as the \textit{rumble} and the \textit{trumpet}~(see \cref{fig:calltypes}).
Each dataset is evaluated at the level at which it is annotated, so the class counts of the two are not directly comparable.

Before use, all recordings are resampled to \qty{16}{\kilo\hertz} and reduced to a single channel.
This is the native input rate of \gls{hubert} and of the \gls{aerd} baseline, and matches the supervised evaluation from which the trained baselines are taken~\autocite{Geldenhuys2026Embed}, so every system is presented with the same band-limited audio.
The resulting \qty{8}{\kilo\hertz} Nyquist limit lies above the frequency ranges reported for the rumble, roar, trumpet and chirp, which together span approximately \qty{10}{\hertz} to \qty{6.2}{\kilo\hertz} and are similar in Asian and African elephants~\autocite{nairVocalizationsWildAsian2009}.
The \ac{ldc} recordings carry a second channel holding spoken field notes, which is discarded, and every segment is peak-normalised to \qty{-3}{\decibel} relative to full scale.

\section{Experimental Setup}
\label{sec:setup}

We follow the order in which the system processes the audio, beginning with segmentation and the embedding models, continuing with the cross-validation and episode construction protocol, and closing with the trained classifiers, the bootstrap and the metrics.

\subsection{Audio segmentation}
\label{sec:segmentation}

Segmentation isolates the intervals of a long recording that correspond to elephant vocalisations, fixing the start and end time of each call.
We assume these endpoints to be known and take the segments from the human annotations.
Because vocalisations may overlap, the task is fundamentally multi-label, and an assigned class counts as correct if it matches any of the human-annotated labels of the segment.

Segmentation of the \ac{elev} recordings yields 1063 segments, of which 594 contain a call and 422 carry a subcall-type label~(see \cref{tab:data}).
Segments shorter than \qty{1}{\second} are discarded.
A call type is evaluated only when it has at least one exemplar in the query fold and one in the support folds of every split.
Call types that cannot meet this requirement are excluded.
This leaves $C$ call types and $Q$ query segments, each classified exactly once.
For \ac{elev}, 12 of the 27 annotated subcall types remain, and for \ac{ldc}, 10 of the 16 annotated call types remain.

\subsection{Embedding models}
\label{sec:emb-model}

All embedding models are fixed throughout, their weights never updated or fine-tuned.
None was trained to distinguish elephant call types, and only \gls{perchv2} encountered elephant recordings during training.

\paragraph{\Acp{mfcc}}
\label{sec:emb:mfcc}
Each segment is represented by the per-frame average of its \ac{mfcc} vectors, computed with \qty{25}{\milli\second} frames, \qty{10}{\milli\second} stride, a 128-band mel filterbank and 40 cepstral coefficients~\autocite{Davis1980MelConfig,hagiwara2022beans}.

\paragraph{\Gls{hubert}}
\label{sec:emb:hubert}
This is a transformer encoder pretrained on English speech with a \ac{ssl} procedure~\autocite{hsu2021hubert}.
We evaluate the \textit{base} variant, which produces a 768-dimensional embedding for each \qty{20}{\milli\second} frame, averaged to give a single representation per segment.
\Textcite{Geldenhuys2026Embed} found the second transformer layer most informative for elephant call classification, so all embeddings are taken from that layer.

\paragraph{\Gls{perchv1}}
\label{sec:emb:perchv1}
This is an \gls{efficientnet-b1} \ac{cnn} trained on avian audio from \textit{Xeno-canto}~\autocite{google2023perch}.
It accepts a fixed five-second input, from which a mel spectrogram with \ac{pcen} is computed using \qty{20}{\milli\second} frames and 160 mel bands, and a 1280-dimensional embedding obtained per frame, the frames being combined by global average pooling.
Our segments on average match this length~(see \cref{tab:data}).
Shorter segments are nonetheless padded, and longer segments are presented as consecutive, non-overlapping \qty{5}{\second} intervals whose embeddings are averaged over the segment.

\paragraph{\Gls{perchv2}}
\label{sec:emb:perchv2}
This model builds on \gls{perchv1} with a larger \gls{efficientnet-b3} \ac{cnn} backbone, which is trained on a multi-taxa corpus of avian, mammalian and environmental audio~\autocite{Merrienboer2025perchv2} rather than avian-only \textit{Xeno-canto} audio, and produces 1536-dimensional embeddings.
Its species label set includes \textit{Loxodonta africana}, \textit{Loxodonta cyclotis} and \textit{Elephas maximus}, therefore its training data include elephant recordings, labelled by species but never by call type.
Its mel spectrogram uses 128 bands and no \ac{pcen}.
In all other respects the embeddings are computed as for \gls{perchv1}.

\subsection{Cross-validated episode construction}
\label{sec:setup:eval}

The $K$-fold cross-validation protocol of \textcite{stone1974crossval,mosteller1968crossval} is adopted unchanged, with five folds for the \ac{elev} dataset and ten for the \ac{ldc} dataset.
Folds are stratified by call type and grouped by recording, so no recording spans folds.
For the purpose of stratification, each segment has a single associated label, determined by the call type at its temporal centre.
However, a classification is counted as correct if it matches any of the annotated labels present in the segment~(see \cref{sec:segmentation}).
For the fully-supervised classifiers~(see \cref{sec:setup:baseline}), each fold serves in turn as the \emph{test fold}, the classifier being trained on the remaining \emph{training folds} together with the common fold.

The centroid classifier reuses these folds, which keeps its results comparable to those of the trained classifiers.
Queries are drawn from the held-out \emph{query fold} and support exemplars from the remaining \emph{support folds}~(see \cref{fig:method}).
Each fold serves as the query fold in turn, so every segment outside the common fold is evaluated exactly once and the reported performance spans the dataset.
Each episode supplies a single query segment~($q = 1$) of~\cref{eq:query} against an independently drawn support set, so the number of episodes $e$ far exceeds $K$.

The centroid classifier is required to face the same decision as the trained classifiers.
Conventional episodic evaluation uses a small fixed number of classes per episode, $N$~(typically 5-way or 10-way), so the classifier need only choose among a sampled subset guaranteed to contain the true class.
We instead set $N = C$, the number of evaluated call types~(see \cref{sec:segmentation}), so every class appears in every episode and the classifier chooses among all $C$ candidates, exactly as the supervised classifiers must.

\subsection{Varying the number of support exemplars}
\label{sec:setup:kshot}

We determine centroid-classification performance as a function of the number of labelled support exemplars per class, $k$.
We increase $k$ from a single exemplar towards the limit at which every eligible support exemplar is used, where the episode reproduces the training regime of the fully-supervised classifiers~(see \cref{sec:setup:baseline}).
Throughout, the \emph{few-shot regime} means $k \le 5$ exemplars per class, following the five-shot convention of the bioacoustic few-shot literature~\autocite{Nolasco2023BioFSL}.

\subsection{Complete and partial support sets}
\label{sec:setup:partial}
Let $a_n$ denote the number of labelled exemplars of class~$n$ in the support folds of a cross-validation split~(see \cref{sec:setup:eval}), while a $k$-shot episode requires $k$ exemplars of every class.
When $k \le a_n$ the support set of the class is \emph{complete}, and when $k > a_n$ it is \emph{partial}, the realised support set size being $\min(k, a_n)$.
An episode is partial once any of its classes is.
We denote by~$k^\star$ the largest $k$ for which every evaluated class is still complete.
The distinction matters for the sampling noise~(see \cref{sec:results}), because a partial class admits only one support set --- its entire eligible pool --- and so no longer contributes to support-sampling variability.

\subsection{Baseline classification systems}
\label{sec:setup:baseline}

A strongly-supervised end-to-end \gls{ast-seq} model, as described in \textcite{geldenhuys2024aerd}, serves as a further baseline and is referred to as the \gls{aerd} system.
It is an \ac{ast} adapted for sequence-to-sequence modelling, pretrained on \gls{audioset}~\autocite{chen2023beats} and trained end-to-end for elephant call classification, detection and endpointing.

The \gls{aerd} system was trained on a reduced set of call types, since several of the types evaluated here occur too infrequently for end-to-end training.
On the \ac{elev} dataset it covers 7 of the 12 evaluated subcall types, and on the \ac{ldc} dataset it groups the annotated call types into five coarser categories, one of which has no counterpart here because its calls fall below the \qty{1}{\second} minimum~(see \cref{sec:segmentation}).
We compare the centroid classifier against this baseline over the reduced set of call types separately in \cref{sec:results}.

\Textcite{Geldenhuys2026Embed} evaluated a range of fixed pretrained embeddings for elephant call classification using several neural classifiers.
We pair \ac{mfcc} features and the \gls{perchv1}, \gls{perchv2} and \gls{hubert-base-l2} embeddings with \acf{lr} and a recurrent classifier, chosen for the range of performance they span in that evaluation.

The \ac{lr} classifier, which we call the \ac{lr} baseline, is linear, so its performance indicates class separability in the embedding space.
For each embedding we also report the performance of a recurrent classifier that was reported to perform well in the supervised experiments reported in~\textcite{Geldenhuys2026Embed}, namely a \gls{gru}, \gls{lstm} or Elman \gls{rnn} according to embedding and dataset~(see \cref{tab:results}).

Within each outer cross-validation fold, the \ac{lr} and recurrent classifiers are trained on \emph{all} labelled data in the training portion of that fold, whereas the centroid classifier sees only the $k$ support exemplars per class.
This asymmetry is intentional, as it ensures that the trained classifiers are strong.
Good centroid performance under controlled $k$ is therefore a strong result, since every trained classifier had access to the full training fold.

\subsection{Bootstrapped estimation of support set sampling noise}
\label{sec:setup:bootstrap}

Classification performance is a random variable over the drawn support set, with the query segments fixed~(see \cref{sec:setup:eval}).
We estimate its distribution with a nonparametric bootstrap~\autocite{Efron1979BootstrapMethods, Efron1994IntroBootstrap}.

Evaluating a query against a support set requires forming the class centroids and scoring the query against them~(see \cref{eq:centroid,eq:centroid:softmax}), so we perform this once, up front, and reuse the resulting scores.
Let $\mathbf{x}_j$, for $j = 1, \ldots, Q$, denote the query segments of the dataset.%
\footnote{We write $Q$ for the total number of query segments, distinct from the per-episode query-set size $q$ of~\cref{eq:query}. Under the $N = C$ protocol of~\cref{sec:setup:eval}, each segment of the query folds is evaluated as a query exactly once, so $Q$ equals the number of query segments reported in~\cref{tab:data}.}
For each query, $R = \num{100}$ support sets are drawn independently from the support folds and scored once up front, giving a $Q \times R \times N$ array of softmax scores.
The $k$ exemplars within a support set are distinct, but an exemplar may recur across the independent draws.

Each of $B = \num{5000}$ bootstrap samples then selects, independently for each query, one of its $R$ precomputed score vectors, and the chosen performance metric~(\ac{map} or \ac{auroc}) is then evaluated over the selected fold, and averaged across folds, as set out in \cref{sec:metrics}.
For each dataset, embedding, and value of $k$ we report the bootstrap mean of these values together with the interval between their 1st and 99th percentiles, whose width is the spread of classification performance, or sampling noise, namely how much the reported performance would vary were the support sets redrawn.
The procedure is metric-agnostic and applies unchanged to any performance summary.

\subsection{Metrics}
\label{sec:metrics}

We report two metrics, the \ac{auroc} and the \ac{map}, both threshold-independent, unlike fixed-threshold measures.

The \ac{roc} curve plots the \ac{tpr} against the \ac{fpr} across thresholds, the area beneath it summarising performance.
It is not biased by the class prior and remains informative under imbalance.
The \ac{auroc} is computed per class and reported as the unweighted mean over call types.

Ranking the query segments by the score for a class ideally places its members highest.
The precision-recall curve traces precision against recall as the threshold descends this ranking, the \ac{ap} of a class is the area beneath it, and the \ac{map} is the unweighted mean of the per-class \ac{ap}.
Both metrics are computed within each cross-validation fold and then averaged across the folds, with each query segment evaluated exactly once in its held-out fold.
The \ac{auroc} counts every correctly ordered pair equally, wherever it falls in the ranking, so the \ac{map} is the more demanding measure for the rare call types.
The \ac{map} is therefore tabulated throughout, with the \ac{auroc} reported in prose where the two disagree.

\section{Results}
\label{sec:results}

We present classification results for the nearest-centroid classifier on the \ac{elev} and \ac{ldc} datasets, using classical \ac{mfcc} features alongside the \gls{perchv1}, \gls{perchv2} and \gls{hubert-base-l2} fixed embeddings.
For each dataset and each embedding we vary the number of labelled exemplars per class, $k$, and compare against the \ac{lr} baseline and the recurrent baseline paired with that embedding~(see \cref{sec:setup:baseline}).
The comparison against the \gls{aerd} baseline requires a reduced set of call types and is therefore presented on its own at the end of this section.
Sampling noise is reported as the 1st-to-99th-percentile spread of the bootstrap distribution~(see \cref{sec:setup:bootstrap}).

\begin{figure}[tb]
	\centering
	\includegraphics[page=1,width=\columnwidth]{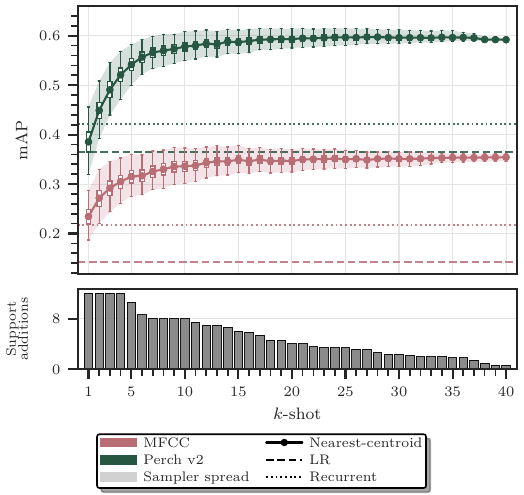}
	\caption{
		Nearest-centroid \ac{map} performance on the \acs{elev} dataset varying $k$, using fixed \ac{mfcc} features and \gls{perchv2} embeddings~($C = 12$).
		Boxes give the interquartile range and median of the bootstrap distribution, whiskers the 1st and 99th percentiles, markers the mean, and the shaded band the bootstrap sampler spread~(see \cref{sec:setup:bootstrap}).
		Dashed and dotted lines mark the \ac{lr} and recurrent baselines, in the colour of the features or embeddings used.
		The lower panel counts the new exemplars per shot~(see \cref{sec:setup:partial}).
	}
	\label{fig:results:kshot:elev}
\end{figure}

\begin{figure}[tb]
	\centering
	\includegraphics[page=1,width=\columnwidth]{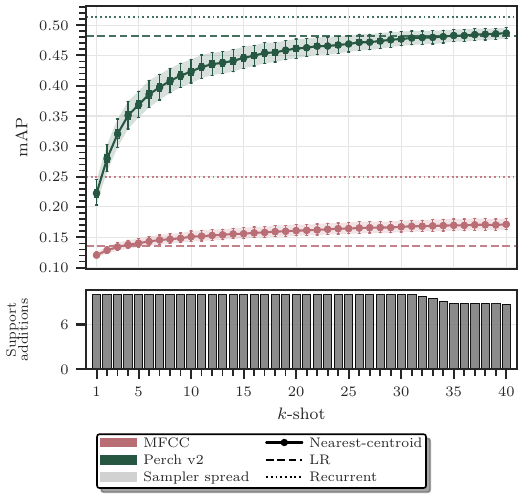}
	\caption{
		Nearest-centroid \ac{map} performance on the \acs{ldc} dataset against $k$, using fixed \ac{mfcc} features and \gls{perchv2} embeddings~($C = 10$).
		Layout and line styles are as in \cref{fig:results:kshot:elev}, with \gls{gru} as both recurrent baselines.
	}
	\label{fig:results:kshot:ldc}
\end{figure}

\Cref{fig:results:kshot:elev,fig:results:kshot:ldc} report centroid-classification performance as a function of $k$ on the \ac{elev} and \ac{ldc} datasets, contrasting \ac{mfcc} features with \gls{perchv2} embeddings, while \cref{tab:results} gives the point estimates for the \ac{mfcc} features and all three embeddings.

From \cref{fig:results:kshot:elev}, centroid-classification performance on the \ac{elev} dataset with \gls{perchv2} embeddings rises with each additional exemplar up to a plateau, the \ac{map} increasing from \num{0.3853} at $k = 1$ to \num{0.5208} at $k = 4$ with the gains between successive values of $k$ diminishing thereafter.
At $k = 1$ the bootstrap estimate already exceeds the \ac{lr} baseline using \gls{perchv2} embeddings~(\ac{map}~$= \num{0.3654}$), which is trained on all labelled data of the training folds.
At $k = 2$ the entire interquartile range of the sampling distribution lies above this baseline, and at $k = 3$ more than \qty{99}{\percent} of the support sets drawn from the data exceed it.
At $k = 2$ the centroid classifier also exceeds, on average, the Elman~\ac{rnn} baseline using \gls{perchv2} embeddings~(\ac{map}~$= \num{0.4220}$), and from $k = 5$ the entire 1st-to-99th-percentile interval lies above that baseline.
Beyond this point performance rises only gradually, approaching a plateau of approximately \num{0.59} at the largest values of $k$ considered~(for $k>20$).
The levelling of the \ac{map} coincides with the decline in new exemplars reported in the lower panel of~\cref{fig:results:kshot:elev}~(see \cref{sec:discussion:plateau}).
With \ac{mfcc} features the same figure shows the classifier following the same shape at a far lower level, its mean remaining below the \ac{lr} baseline using \gls{perchv2} embeddings at every value of $k$ considered.

The sampling distributions are approximately symmetric at every $k$, with the bootstrap mean and the median almost coincident, and they narrow as $k$ increases.
This narrowing follows from the limited labelled data rather than from the classifier, since a class that has exhausted its eligible exemplars supplies the same support set in every episode and so contributes no variability~(see \cref{sec:setup:partial}).
Clearance of a trained classifier by the entire 1st-to-99th-percentile interval beyond $k^\star = 4$ is therefore not in itself evidence of good generalisation.

\begin{table*}[t]
	\centering
	\caption{
		Performance of the $k$-shot centroid classifier on the \ac{elev} and \ac{ldc} datasets as $k$ is varied, reported as \ac{map}.
		The two rows at the bottom of each subtable gives the classification performance of the \ac{lr} baseline and the best recurrent baseline~(see \cref{sec:setup:baseline}): Elman \ac{rnn} for \gls{perchv2} on \acs{elev}, \gls{lstm} for \gls{perchv1} on \acs{ldc}, \gls{gru} otherwise, none for \gls{hubert-base-l2}.
		Every class remains complete up to $k^\star = 4$ on the \acs{elev} dataset and $k^\star = 31$ on the \acs{ldc} dataset, and rows beyond these values draw from partial support sets~(see \cref{sec:setup:partial}).
		A star marks the first tabulated value of $k$ that draws on partial support sets.
		A single underline indicates the first value surpassing \acs{lr}.
		A double underline indicates the first value to surpass the recurrent baseline, which also exceeds \acs{lr}.
	}
	\label{tab:results}
	\small
	\sisetup{table-format=1.4, round-precision=4, round-pad=true}
	\setlength{\tabcolsep}{2pt}
	\begin{subtable}[t]{0.49\textwidth}
		\centering
		\caption{\acs{elev}}
		\label{tab:results:elev}
		\begin{tabular}{@{}rSSSS@{}}
			\toprule
			$k$ & {\acs{mfcc}} & {\gls{perchv1}} & {\gls{perchv2}} & {\gls{hubert-base-l2}} \\ \midrule
			1  & {\passrnn{0.2344}} & 0.3159             & {\passlr{0.3853}}  & {\passlr{0.3288}} \\
			2  & 0.2716             & {\passrnn{0.4157}} & {\passrnn{0.4491}} & 0.3935            \\
			3  & 0.2919             & 0.4602             & 0.4907             & 0.4370            \\
			4  & 0.3049             & 0.4968             & 0.5208             & 0.4584            \\
			$^\star5$  & 0.3152             & 0.5121             & 0.5415             & 0.4715            \\
			10 & 0.3361             & 0.5505             & 0.5781             & 0.5011            \\
			40 & 0.3542             & 0.5898             & 0.5920             & 0.5171            \\
			\midrule
			LR  & 0.1418 & 0.3628 & 0.3654 & 0.3068 \\
			RNN & 0.2177 & 0.3843 & 0.4220 & {--}   \\ \bottomrule
		\end{tabular}
	\end{subtable}
	\hfill
	\begin{subtable}[t]{0.49\textwidth}
		\centering
		\caption{\acs{ldc}}
		\label{tab:results:ldc}
		\begin{tabular}{@{}rSSSS@{}}
			\toprule
			$k$ & {\acs{mfcc}} & {\gls{perchv1}} & {\gls{perchv2}} & {\gls{hubert-base-l2}} \\ \midrule
			1  & 0.1206            & 0.1704 & 0.2226            & 0.1718 \\
			2  & 0.1285            & 0.2232 & 0.2799            & 0.2037 \\
			3  & 0.1340            & 0.2596 & 0.3204            & 0.2264 \\
			4  & {\passlr{0.1378}} & 0.2838 & 0.3509            & 0.2398 \\
			5  & 0.1402            & 0.3017 & 0.3684            & 0.2492 \\
			10 & 0.1510            & 0.3507 & 0.4233            & 0.2819 \\
			$^\star40$ & 0.1713            & 0.4246 & {\passlr{0.4868}} & 0.3326 \\
			\midrule
			LR  & 0.1359 & 0.4384 & 0.4824 & 0.3492 \\
			RNN & 0.2489 & 0.4411 & 0.5126 & {--}   \\ \bottomrule
		\end{tabular}
	\end{subtable}
\end{table*}

The corresponding evolution of \ac{map} for the \ac{ldc} dataset is shown in \cref{fig:results:kshot:ldc} and follows broadly the same trend, with three differences, again for \gls{perchv2} embeddings.
First, the centroid classifier remains below the recurrent baseline at every value of $k$ considered, and reaches the \ac{lr} baseline using \gls{perchv2} embeddings~(\ac{map}~$= \num{0.4824}$) only at the large value of $k = 40$, having become level near $k = 35$.
The stronger \gls{gru} baseline using \gls{perchv2} embeddings~(\ac{map}~$= \num{0.5126}$) is never matched.
Second, no class exhausts its eligible pool until $k^\star = 31$, far beyond the few-shot regime.
Third, the spread of the classifier performance as a result of support set composition remains narrow and approximately constant as $k$ increases, rather than narrowing as it does on the \ac{elev} dataset.
The larger \ac{ldc} pool of training examples means that every class remains complete well beyond the few-shot regime, and so the support sets available do not contract and the spread does not narrow.
We interpret the first two differences in~\cref{sec:discussion:plateau}.

The \ac{map} point estimates for the \ac{mfcc} features and all three embeddings are reported in \cref{tab:results}.
The ordering of the embedding performance is largely consistent across both datasets, with \gls{perchv2} strongest, followed by \gls{perchv1}, \gls{hubert-base-l2}, and \ac{mfcc}.
The crossover behaviour described above for \gls{perchv2} extends to the other embeddings.
On the \ac{elev} dataset~(see \cref{tab:results:elev}), the centroid classifier exceeds the corresponding \ac{lr} baseline with every feature set and embedding at the larger values of $k$ considered, and exceeds the corresponding recurrent baseline in each of the three cases where one is reported.
On the \ac{ldc} dataset~(see \cref{tab:results:ldc}) only the \ac{lr} baseline is overtaken, with \ac{mfcc} features from $k = 4$ and with \gls{perchv2} embeddings at $k = 40$, and no embedding reaches the corresponding recurrent baseline.
In terms of \ac{auroc} this ordering does not hold at every value of $k$ considered, and two of these crossings are absent, the centroid classifier staying below the \ac{lr} baseline with \gls{hubert-base-l2} embeddings and below the recurrent baseline with \gls{perchv2}.

The comparison against the \gls{aerd} baseline is presented separately, because it requires the centroid classifier to be scored over the reduced set of call types on which that baseline was trained~(see \cref{sec:setup:baseline}), so that the two are directly comparable.
We also restrict the comparison to the strongest two embeddings, \gls{perchv1} and \gls{perchv2}.
The centroid classifier still assigns each query among all $C$ classes, so restricting the scoring does not make its task easier.
However, these values are not comparable with those of \cref{tab:results}, which score every evaluated call type.
On the \ac{elev} dataset, restricted to the 7 subcall types --- used by the \gls{aerd} baseline --- the centroid classifier with \gls{perchv2} embeddings attains an \ac{map} of \num{0.3926} at $k = 1$ and \num{0.6011} at $k = 40$, compared with \num{0.4054} for that baseline~\autocite{geldenhuys2024aerd}.
Its mean crosses the \gls{aerd} baseline at $k = 2$, and from $k = 3$ the entire 1st-to-99th-percentile interval lies above it.
For the \gls{perchv1} embeddings, it crosses at $k = 5$~(\ac{map}~$= \num{0.4672}$).
In terms of \ac{auroc}, neither embedding reaches the \gls{aerd} baseline at any evaluated $k$, \gls{perchv2} attaining \num{0.8333} at $k = 40$ against \num{0.8453}.
On the \ac{ldc} dataset, 8 of the 10 evaluated call types fall under the four \gls{aerd} categories that have a counterpart here~(see \cref{sec:setup:baseline}), namely the rumble, the roar, the bark and the trumpet, the remaining two being the non-vocal \textit{blow} and \textit{trunk-bounce}.
Scored over those constituent calls, the centroid classifier using \gls{perchv2} embeddings attains an \ac{map} of \num{0.2705} at $k = 1$ and \num{0.5285} at $k = 40$, so it remains below the \gls{aerd} baseline~(\ac{map}~$= \num{0.6412}$) at every evaluated $k$.
Its shortfall at $k = 40$ is smaller in terms of \ac{auroc}~(\num{0.8838} against \num{0.9497}) than \ac{map}.
The parameter-free centroid classifier therefore overtakes the strongly-supervised \gls{aerd} baseline on the data-scarce \ac{elev} dataset in \ac{map} alone, and on the larger \ac{ldc} dataset in neither metric.

\section{Discussion}
\label{sec:discussion}

We interpret the results of~\cref{sec:results} in two ways.
First, we view the nearest-centroid rule as a linear classifier.
Second, we consider the performance plateau of the nearest-centroid classifier observed as $k$ is increased.
Together, these arguments clarify when the parameter-free nearest-centroid classifier is preferable to a trained classifier using the same fixed embedding.

\subsection{The nearest-centroid classifier is a linear classifier}
\label{sec:discussion:linear}

\Cref{eq:centroid:linear} shows the nearest-centroid rule to be a linear classifier whose weights and biases are determined entirely by the class centroids.
\Textcite{Snell2017Protonets} make the same observation for prototypical networks under squared Euclidean distance.
It is therefore not a different type of model from the trained \ac{lr} baseline, but has the same linear form with parameters set rather than \emph{fitted}.
Taking this view, \cref{sec:results} compares two ways of fixing those parameters, from a handful of exemplars or by optimisation against all labelled data of the fold.

The two strategies are at opposite ends of the bias-variance trade-off.
Setting the weights using the class means imposes a strong inductive bias, namely an isotropic within-class covariance, which removes the estimation variance an optimiser incurs at the cost of adaptability.
Fitting rather then setting the weights removes that bias, but reintroduces the variance.
This is harmful when labelled exemplars are few, so the trade favours the centroid classifier in that regime.
Gradient-based refinement would recover the trained classifier given sufficient data.
The value of doing so is precisely what the two datasets probe.

\subsection{The performance plateau}
\label{sec:discussion:plateau}

On the \ac{elev} dataset, centroid-classification performance rises with each additional exemplar but with diminishing returns, reaching a plateau~(see \cref{fig:results:kshot:elev}).
The \emph{shape} of this rise is common to all four embeddings, whereas the \emph{level} at which it settles is not, reaching \ac{map}~$\approx \num{0.59}$ with \gls{perchv1} and \gls{perchv2} embeddings but only \num{0.52} with \gls{hubert-base-l2} embeddings and \num{0.35} with \ac{mfcc} features.
Two mechanisms govern the shape, the first tracing the diminishing returns to the embedding and the second to the exhaustion of the labelled pool.

The first mechanism operates within each class.
Each centroid is the sample mean of its support embeddings~(see \cref{eq:centroid}), so as $k$ grows it converges towards the class-conditional mean, the variance of the estimate falling as roughly $1/k$ and each successive exemplar displacing it less than the last.
The class representation is then fixed by the geometry of the embedding space.
However, convergence of the mean does not guarantee that the mean is a faithful summary of the class.
A multimodal class cannot be represented by a single centroid, and further exemplars only sharpen the estimate of an unrepresentative location between the modes.
This is the isotropic within-class assumption noted in~\cref{sec:discussion:linear}, and nothing in the pretraining of the embedding models encourages it to hold.

The second mechanism operates across classes and explains why the gains cease rather than merely slow.
Classes differ in the number of eligible exemplars $a_n$ in their support folds and so become partial at different rates.
Beyond $k^\star = 4$ on the \ac{elev} dataset the smaller classes are partial, their centroids fixed at the mean of the entire eligible pool~(see \cref{sec:setup:partial}).
Because the \ac{map} is the unweighted mean of the per-class \ac{ap}, gains can then arise only through the classes that remain complete, so once those are exhausted the curve plateaus.

On the \ac{ldc} dataset every class remains complete up to $k^\star = 31$, so the second mechanism comes only well beyond the few-shot regime.

Whichever mechanism halts the rise, the embedding fixes the \emph{level} at which the plateau settles.
The centroid rule places the burden of discrimination on the fixed representation~(see \cref{sec:background:centroid}), so once the centroids converge the residual error is the class overlap in that space, which further exemplars cannot reduce.
That both Perch models settle above \gls{hubert-base-l2} is consistent with reports that embeddings from large-scale bird classifiers transfer well to non-avian taxa~\autocite{google2023perch}, and that \gls{perchv2} improves on \gls{perchv1} in few-shot whale call classification~\autocite{Burns2025Perch2Whales}, as it does here at small $k$ and on the \ac{ldc} dataset.
Over the reduced set of call types on which the \gls{aerd} baseline was trained~(see \cref{sec:results}), the centroid classifier reaches a plateau that is superior in terms of \ac{map} with \gls{perchv1} and \gls{perchv2} embeddings, even though the feature extractor of that baseline is trained end-to-end on the elephant-call task.
With the weaker \ac{mfcc} and \gls{hubert-base-l2} representations, the plateau is lower, and cannot be raised by a larger pool of support data.

This matters most for call types that are absent from the pretraining data, since where the embedding does not encode features that separate such a class, no quantity of support exemplars makes its pooled prototype discriminative.

For the \ac{elev} dataset, the plateau in nearest-centroid classification performance lies clearly above both the \ac{lr} baseline and the Elman~\ac{rnn} baseline, whereas for the \ac{ldc} dataset it remains below the corresponding \ac{lr} baseline and \ac{gru} baseline, a reversal that can be ascribed to the bias-variance trade-off discussed in \cref{sec:discussion:linear}.
The \ac{ldc} dataset comprises more than fifteen times as many query segments as the \ac{elev} dataset.
No class becomes partial until $k^\star = 31$, so every call type retains enough exemplars to fit the trained classifier, which exploits statistical structure the centroid classifier discards by design.

On the \ac{ldc} dataset the centroid classifier needs all 40 exemplars per class merely to draw level with the \ac{lr} baseline using \gls{perchv2} embeddings, and with \gls{perchv1} embeddings it does not reach that baseline at any value of $k$ considered~(see \cref{tab:results:ldc}).
This shortfall measures what the parameter-free rule forgoes, namely adaptation to the distribution of each class rather than reliance on the isotropic assumption.
The \ac{lr} and recurrent baselines are themselves close on the stronger embeddings~(see \cref{tab:results:ldc}), so the shortfall stems from that bias and not from any lack of classifier capacity.

\section{Summary and Conclusion}
\label{sec:conclusion}
We have presented a comparative evaluation of parameter-free episodic nearest-centroid classification of elephant vocalisations on fixed pretrained acoustic embeddings.
Experiments were performed on the \ac{elev} and \ac{ldc} datasets, using the \gls{perchv1}, \gls{perchv2} and \gls{hubert-base-l2} embeddings alongside \ac{mfcc} features.
Evaluation was performed in an $N$-way $k$-shot manner and under the same stratified $K$-fold cross-validation protocol used by the conventionally trained supervised baselines.
A nonparametric bootstrap over the $R = \num{100}$ support sets drawn independently for each query quantified the sampling noise of the resulting classification performance.

On the low-resource \ac{elev} dataset, the parameter-free centroid classifier proved to be notably data-efficient.
Using the stronger \gls{perchv1} and \gls{perchv2} embeddings, it surpassed the fully-trained recurrent baseline in terms of \ac{map} from two exemplars per class.
Compared over the reduced set of call types --- on which the strongly-supervised end-to-end \gls{aerd} baseline was trained --- the centroid classifier using \gls{perchv2} embeddings overtook that baseline in \ac{map} as well, again from two exemplars per class.
This is a strong result, despite the absence of any learnable parameters.

On the larger \ac{ldc} dataset, the recurrent and end-to-end baselines retained their advantage for all considered values of $k$, as expected.
Only the weaker \ac{lr} baseline was overtaken, in \ac{map} alone and with \gls{perchv2} embeddings, at $k = 40$ and therefore far outside the few-shot regime.

Together the two regimes delineate when the parameter-free classifier is preferable, namely when labelled exemplars are few and the embedding already encodes the separating features.
The centroid classifier is therefore most advantageous where labelled data are scarce, and its advantage diminishes as they become abundant.
For a conservation practitioner the implication is direct.
Given a handful of labelled exemplars per call type and a suitable pretrained embedding, a centroid classifier can be deployed without the training, validation and tuning stages a trained classifier demands.
That is the setting in which scarce-label tasks such as subcall classification most often arise.
The remaining limit is then the embedding itself, none of those evaluated having been trained to distinguish elephant call types, so extending pretraining to such call types and other taxa is where further gains are most likely to lie.

\section*{Acknowledgements}
\begin{hidden}[\textit{Acknowledgements omitted for purpose of blind peer-review}]
We gratefully acknowledge financial support from Telkom (South Africa) for the research presented in this paper.
We also thank the Stellenbosch Rhasatsha High Performance Computing facility and team for access to their facilities and for their technical support, which were invaluable to this project.
Finally, we thank the anonymous reviewers for the time and insights they gave towards this manuscript.
\end{hidden}

\section*{ORCID}
\begin{hidden}[\textit{ORCID identifiers omitted for purpose of blind peer-review}]
\noindent\small
\begin{tabular*}{\linewidth}{@{}l@{\extracolsep{\fill}}r@{}}
	\textit{Christiaan M. Geldenhuys} &
		\href{https://orcid.org/0000-0003-0691-0235}%
		     {\includegraphics[width=1em]{orcid.pdf}\hspace*{0.4em}orcid.org/0000-0003-0691-0235} \\
	\textit{Thomas R. Niesler} &
		\href{https://orcid.org/0000-0002-7341-1017}%
		     {\includegraphics[width=1em]{orcid.pdf}\hspace*{0.4em}orcid.org/0000-0002-7341-1017} \\
\end{tabular*}
\end{hidden}

\printbibliography

\end{document}